\documentclass[twocolumn,superscriptaddress,showpacs,preprintnumbers,amsmath,amssymb,prl,reprint]{revtex4}
\usepackage{graphicx}
\usepackage{dcolumn,xcolor,ulem}
\usepackage{amsmath} 
\usepackage{amssymb}
\usepackage{amsfonts}
\usepackage{dsfont}
\usepackage{bm}
\usepackage[latin1]{inputenc}
\usepackage{hyperref}
\usepackage{placeins}
\usepackage{soul}

\begin{document}

 \title{Nonlinear viscoelasticity and generalized failure criterion for polymer gels}

\author{Bavand Keshavarz}
\email{bavand@mit.edu}
\affiliation{Department of Mechanical Engineering, Massachusetts Institute of Technology, \\ 77 Massachusetts Avenue, Cambridge, Massachusetts 02139, USA}
\author{Thibaut Divoux}
\affiliation{Centre de Recherche Paul Pascal, CNRS UPR 8641 - 115 avenue Schweitzer, 33600 Pessac, France}
\affiliation{MultiScale Material Science for Energy and Environment, UMI 3466, CNRS-MIT, 77 Massachusetts Avenue, Cambridge, Massachusetts 02139, USA}
\author{S\'ebastien Manneville}
\affiliation{Univ Lyon, Ens de Lyon, Univ Claude Bernard, CNRS, Laboratoire de
Physique, F-69342 Lyon, France}
\author{Gareth H. McKinley}
\affiliation{Department of Mechanical Engineering, Massachusetts Institute of Technology, \\ 77 Massachusetts Avenue, Cambridge, Massachusetts 02139, USA}

\date{\today}

\begin{abstract}
Polymer gels behave as soft viscoelastic solids and exhibit a generic nonlinear mechanical response characterized by pronounced stiffening prior to irreversible failure, most often through macroscopic fractures. Here, we aim at capturing the latter scenario for a protein gel using a nonlinear integral constitutive equation built upon ($i$) the linear viscoelastic response of the gel, here well described by a power-law relaxation modulus, and ($ii$) the nonlinear viscoelastic properties of the gel, encoded into a ``damping function". Such formalism predicts quantitatively the gel mechanical response to a shear start-up experiment, up to the onset of macroscopic failure. Moreover, as the gel failure involves the irreversible growth of macroscopic cracks, we couple the latter stress response with Bailey's durability criterion for brittle solids in order to predict the critical values of the stress $\sigma_c$ and strain $\gamma_c$ at the failure point, and how they scale with the applied shear rate. The excellent agreement between theory and experiments suggests that the crack growth in this soft viscoelastic gel is a Markovian process, and that Baileys' criterion extends well beyond hard materials such as metals, glasses, or minerals. 
\end{abstract}

\maketitle

\textit{Introduction.-} Polymer gels find ubiquitous applications in material science, from biological tissues to manufactured goods, among which food stuffs and medical products are the most widespread \cite{Mezzenga:2005,Thiele:2014,Webber:2016}. These materials commonly feature a porous microstructure filled with water, which results in solid-like viscoelastic mechanical properties. While soft polymer gels share common features with hard materials, including delayed failure \cite{Bonn:1998,Leocmach:2014}, crack propagation \cite{Daniels:2007,Baumberger:2006} or work-hardening \cite{Schmoller:2010}, their porous microstructure also confers upon them remarkable nonlinear viscoelastic properties. Indeed, such soft solids strongly stiffen upon increasing deformation, which stems from the inherent nonlinear elastic behavior of the polymer chains composing the gel network \cite{Storm:2005,Piechocka:2010,Carrillo:2013,DeCagny2016a}. Polymer gels hence endure large strains to failure and dissipate substantial mechanical work, leading to very tough hydrogels and elastomers \cite{Zhao:2014}. However, to date no quantitative link has been made between the nonlinear viscoelasticity of polymer gels and the failure that is subsequently observed as the strain-loading is increased beyond the initial stiffening regime.

In the present Letter, we apply the concept of a strain damping function, traditionally used for polymeric liquids and rubber-like materials \cite{Rolon:2009}, to quantify the nonlinear viscoelastic response of a prototypical protein gel. The form of the damping function is constructed experimentally through a series of independent stress relaxation tests that allow us to probe large deformations while injecting very little energy into the gel, hence limiting as much as possible any plastic damage. Following the Boltzmann superposition principle, the damping function is used to construct a time-strain separable constitutive equation of K-BKZ (Kaye--Bernstein-Kearsley-Zapas) form \cite{Bird:1987,Larson:1999} that predicts the gel mechanical response to steady-shear experiments. This approach robustly captures the strain-stiffening of the gel during start up of steady shear tests up to the appearance of a stress maximum that is  accompanied by the onset of the first macroscopic crack. Moreover, in order to link the nonlinear viscoelastic response of the gel to its subsequent brittle-like rupture, we adopt the Bailey criterion, which describes the gel failure as arising from accumulation of irreversible damage \cite{Bartenev:1968,Freed:2002}. The combination of the stress response predicted by the K-BKZ constitutive formulation with the Bailey criterion allows us to predict accurately the scaling of the critical stress and strain at failure with variations in the magnitude of the applied shear rate. Our results extend Bailey's criterion to viscoelastic soft solids and provide a unified consistent framework to describe the failure of protein gels under various shear loading histories.

\textit{Experimental.-} We consider two acid-induced protein gels with substantially different mechanical properties: the first one shows pronounced strain-hardening, while the second does not. They are prepared by dissolving caseinate powder (Firmenich) at 4\% wt. (resp. 8\% wt.) in deionized water under gentle mixing at 600~rpm and $T=35^{\circ}$C. Homogeneous gelation is induced by dissolving 1\% wt. (resp. 8\% wt.) glucono-$\delta$-lactone (GDL, Firmenich) in the protein solution \cite{Lucey:1997,OKennedy:2006}. While still liquid, the protein solution is poured into the gap of a cylindrical Couette shear cell connected to a strain-controlled rheometer (ARES, TA Instruments, Delaware) \cite{remark0}. In situ gelation is achieved within 12~hours after which either a step strain or a constant shear rate is imposed on the sample while the resulting stress response is monitored. In both cases, images of the gel deformation are recorded simultaneously to rheology in order to monitor the nucleation and growth of cracks.\\ 
\begin{figure}[t]
\vspace{-1.85mm}
\centering
\includegraphics[width=0.95\columnwidth]{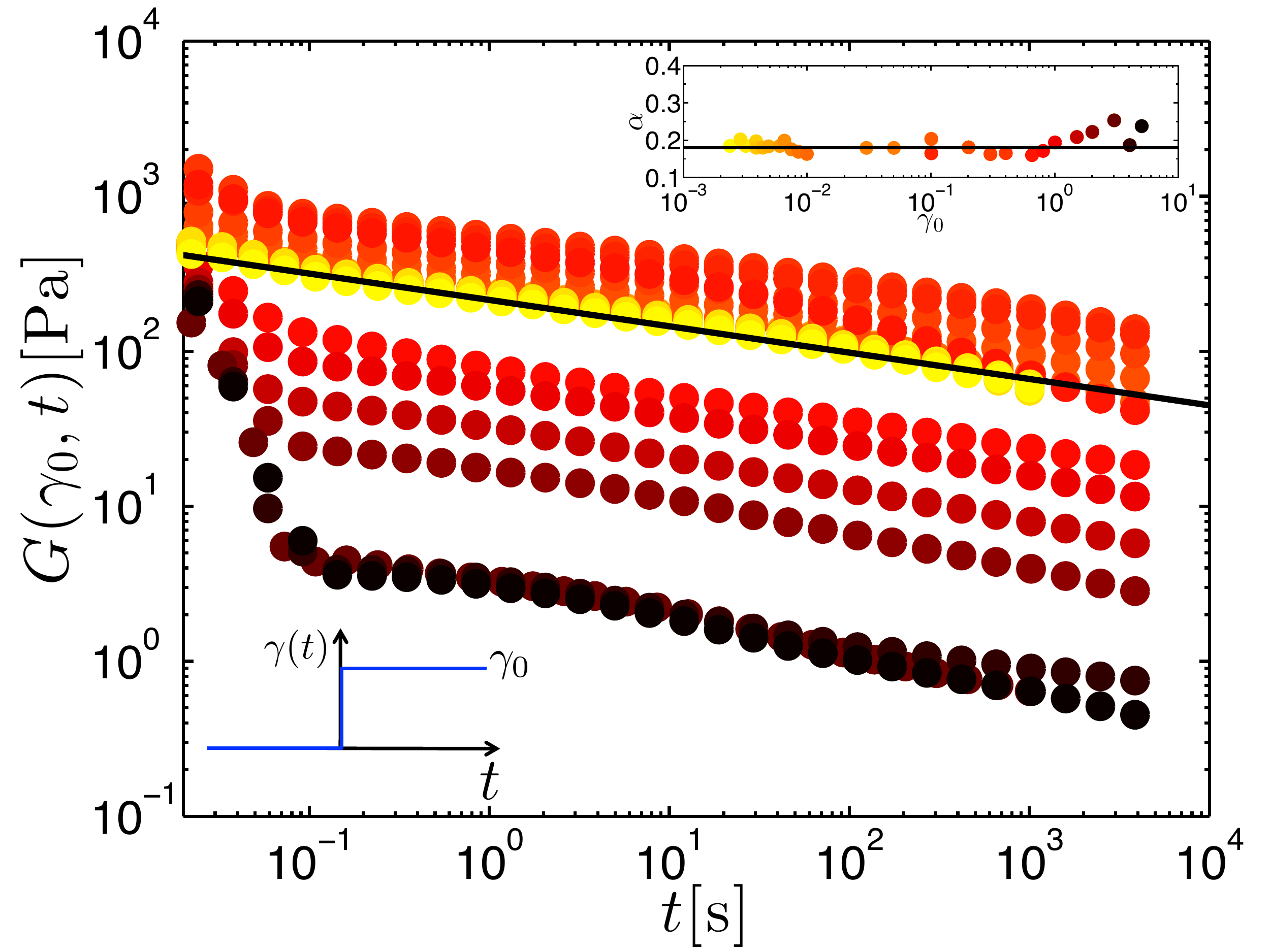}\vspace{-2mm}
\caption{(color online) Nonlinear relaxation function $G(\gamma_0,t)=\sigma(t)/\gamma_0$ vs time $t$ determined by step strain tests, each one performed on a freshly prepared 4\% wt. casein gel. Colors from blue to red  represent strain values ranging from $\gamma_0=0.002$ to $\gamma_0=5$. The black line is the best power-law fit of the data in the linear region ($\gamma_0 \leq 0.01$). Inset: Stress relaxation exponent $\alpha$ extracted from the power-law fit of the data shown in the main graph with the same color code. The horizontal line is the average exponent $\alpha=0.18\pm 0.01$. 
\label{fig1}\vspace{-4mm}}
\end{figure} 
\textit{Damping function.-} To first characterize the viscoelastic properties of the 4\% wt. casein gel, we perform a series of step strain tests. Each experiment is performed on a freshly prepared gel and consists of two successive strain steps. The first step is applied within the linear deformation regime and the stress relaxation is followed over the next 4000~s and serves as a reference for the comparison of two independent experiments. This is followed by a second step at a strain amplitude chosen between $10^{-3}\leq \gamma_0\leq 5$ and the stress is monitored again for 4000~s to measure the gel viscoelastic response. The stress relaxation functions $G(\gamma_0,t)=\sigma(t)/\gamma_0$ associated with the second step of strain are reported in Fig.~\ref{fig1}. At low applied strains ($\gamma_0 \lesssim 0.01$), the magnitude of the viscoelastic stress scales linearly with the imposed strain and the relaxation modulus exhibits a remarkable power-law decrease over four decades of time, which is well modeled by a spring-pot (or fractional viscoelastic element) \cite{Jaishankar:2013}, $G(t) = \mathds{V}t^{-\alpha}/\Gamma(1-\alpha)$, where $\mathds{V}$ and $\alpha$ are the only two material properties required to characterize the gel, and $\Gamma$ denotes the Gamma function. By fitting the data for $\gamma_0\leq0.01$ we find that the relaxation exponent $\alpha =0.18\pm 0.01$ and the prefactor or ``quasiproperty"  $\mathds{V}=266\pm 5$~Pa.s$^{\alpha}$ \cite{Jaishankar:2013,remark1a}. For $\gamma_0 \gtrsim 0.01$, the stress relaxation still exhibits a power-law decrease in time, with the same exponent $\alpha$, after $t \gtrsim 0.1$~s but the magnitude of the stress at a given time first stiffens and then softens as $\gamma_0$ is increased.

\begin{figure}[t!]
\centering
\includegraphics[trim = 0mm 03mm 0mm 0mm, clip,width=0.9\columnwidth]{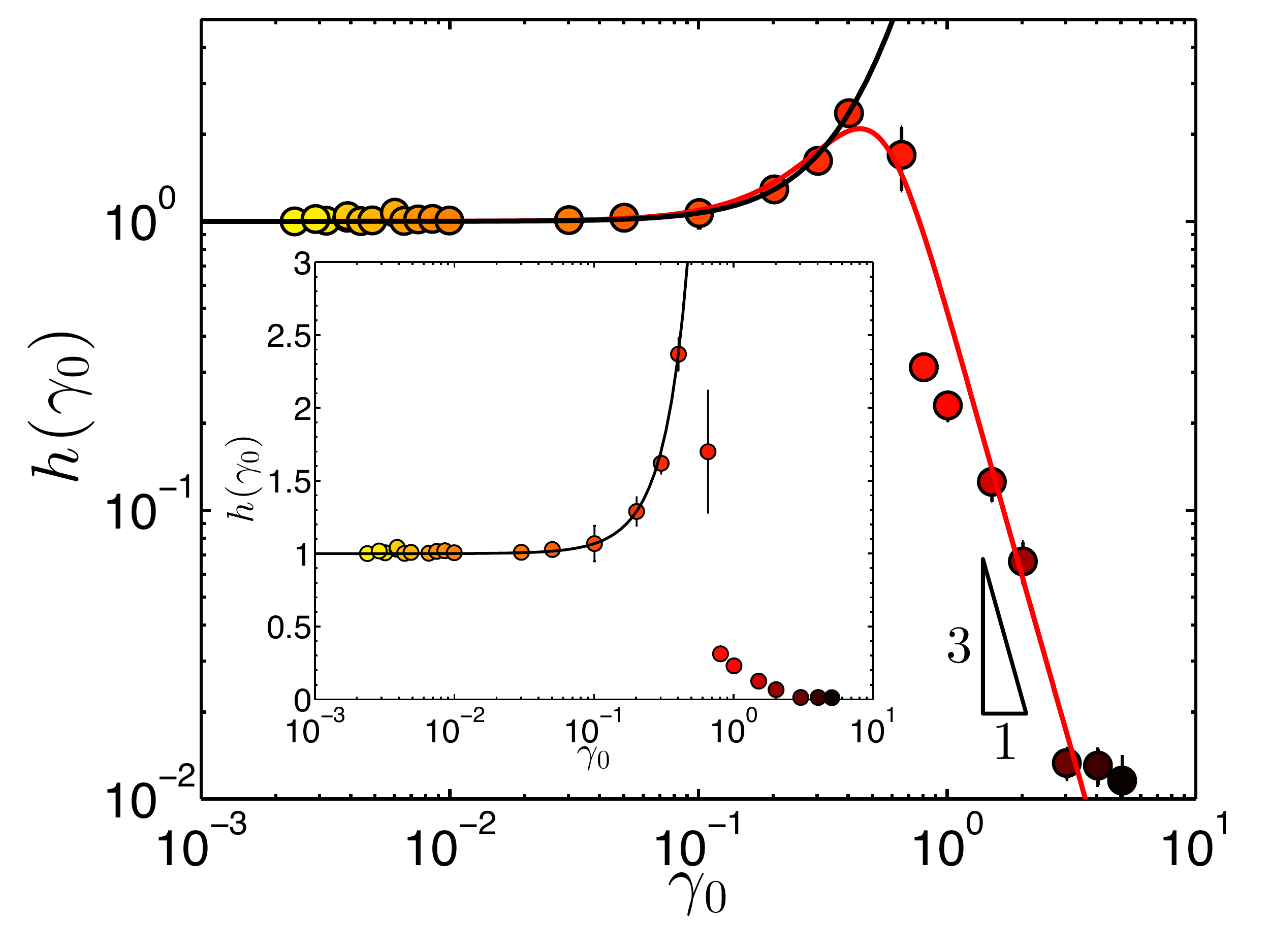}\vspace{-2mm}
\caption{(color online) Strain damping function $h(\gamma_0)$ of a 4\% wt. casein gel as defined in the text. 
Same color code as in Fig.~\ref{fig1}. The solid black line is the best fit function by power series in $\gamma_0^2$ as proposed in \cite{Gisler:1999} which captures the stiffening behavior, but does not account for the softening part of the gel response at strains larger than 50\%. The red continuous line is the best fit function $\tilde h(\gamma)$ of the data (see text). Inset: same data plotted in semilogarithmic scales.
\label{fig2}\vspace{-4mm}}
\end{figure} 
\begin{figure*}[ht!]
\centering
\includegraphics[trim={0 101 0 105},clip,width=0.9\linewidth]{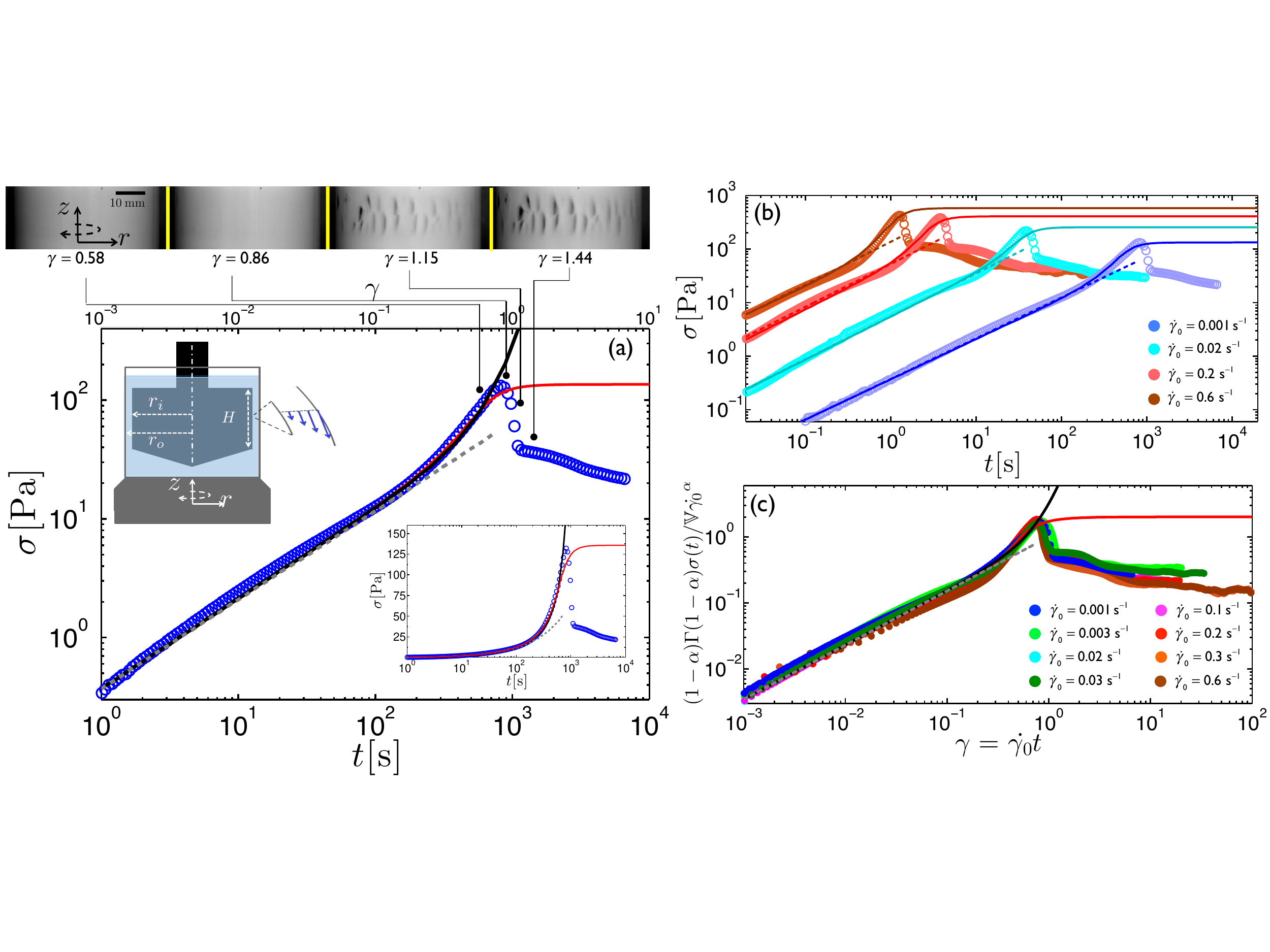}\vspace{-2.0mm}
\caption{(color online) (a) Stress response $\sigma$ vs time $t$ (lower axis) and vs strain $\gamma=\dot\gamma_0 t$ (upper axis) of a 4\% wt. casein gel to a constant shear rate $\dot \gamma_0 =10^{-3}$~s$^{-1}$ initiated at $t=0$. The gray dashed line corresponds to linear viscoelastic response [Eq.~(\ref{eq1})]. The black line corresponds to the K-BKZ equation constructed using only the strain-hardening part of the damping function, $h^*(\gamma_0)$ and reported as the solid black line in Fig.~\ref{fig2}. The continuous red line corresponds to the K-BKZ equation built upon $\tilde h$, which includes both the hardening and the softening components of the damping function. Lower inset: same data on semilogarithmic scales. Upper inset: sketch and images of the side view of Couette cell at different strains recorded simultaneously to the experiment reported in the main graph. (b) Stress responses to individual constant shear rate experiments ranging from $10^{-3}$~s$^{-1}$ to 0.6~s$^{-1}$. Dashed lines indicate the linear response [Eq.~(\ref{eq1})], and the continuous lines correspond to the K-BKZ predictions using $\tilde h$ [Eq.~(\ref{eq2})]. (c) Normalized stress responses $(1-\alpha)\Gamma(1-\alpha)\sigma(t)/\mathds{V}\dot \gamma_0^{\alpha}$ vs strain $\gamma$ for all constant shear rate experiments. 
\label{fig3}\vspace{-4mm}}
\end{figure*} 

Since $\alpha$ is insensitive to the strain amplitude, we can use the concept of strain-time separability \cite{Rivlin:1971} to quantify the strain dependence of the stress relaxation response by computing the  \textit{damping function} \cite{Rolon:2009}, defined as $h(\gamma_0) = \langle G(\gamma_0,t)/G(t)\rangle_t$ where $\langle...\rangle_t$ denotes the time average for $1\leq t\leq 1000$~s for each of the step-strain experiments. The resulting damping function reported in Fig.~\ref{fig2} thus fully characterizes the strain dependence of the viscoelastic response in the material. The gel displays a linear response (i.e. $h=1$) up to $\gamma_0 =0.1$, whereas for intermediate strain amplitudes, the gel exhibits a pronounced strain-stiffening, that is characterized by a maximum value $h \simeq 2.2$ reached at $\gamma=0.5$. Finally, for even larger strains, the material softens due to network rupture and the damping function decreases abruptly as a power-law function of the imposed strain with an exponent of $\sim-3$. Yet, we emphasize that for all step strain tests, the gel remains visually intact even at strain amplitudes as large as $\gamma_0\simeq 5$ \cite{footnote}.    

As proposed in \cite{Gisler:1999,Pouzot:2006}, the strain hardening portion of the damping function is captured in a power series expansion of $h^\star (\gamma_0)$ \cite{remark2}, with the fractal dimension $d_b$ of the stress-bearing network backbone as the only fitting parameter. Here we find $d_b=1.3\pm 0.1$ in good agreement with other measurements for polymer gels (see black line in Fig.~\ref{fig2})\cite{Mellema:2002}. However, describing the whole damping function also requires us to take into account the gel softening that is measured at strains larger than 0.5. Following analogous approaches in the literature for rubbery networks and polymer melts \cite{Rolon:2009} we use the following functional form $\tilde{h}(\gamma_0)=[1+(\gamma_0/\gamma_m)^2]/[1+\left|\gamma_0/\gamma_M\right|^5]$, where $\gamma_m=0.34$ and $\gamma_M=0.57$ are fitting parameters that respectively mark the departure from linearity and the location of the strain maximum. The quadratic dependence of the numerator is set by symmetry \cite{Gisler:1999} while the exponent in the denominator, which depends on the strength of the individual network bonds in the gel, is fixed by a fit of the damping function at large strains $\gamma \gg \gamma_M$. This form describes the whole data set remarkably well (Fig.~\ref{fig2}) and can be used to predict the gel viscoelastic response to start up of steady shear all the way to sample failure as we will now illustrate.

\textit{K-BKZ description of shear start-up.-} In Fig.~\ref{fig3}, we show the evolution of stress and onset of cracking when a 4\% wt. casein gel is submitted to a constant shear rate $\dot \gamma_0=$10$^{-3}$~s$^{-1}$. The stress growth $\sigma(\gamma$) can be separated into three consecutive regimes: a linear viscoelastic regime, characterized by a power-law growth of $\sigma(t)$ up to $\gamma=\dot{\gamma}_0t \simeq 0.2$, followed by a strain-stiffening regime in which $\sigma$ shows a steeper increase up to a critical strain $\gamma_c\simeq 0.8$ at which the stress goes through a maximum $\sigma_c$. Finally, in a third regime, the stress exhibits an abrupt decrease followed by a slower relaxation at larger strains. The gel remains visually intact and homogeneous initially, and the first macroscopic fracture appears at the end of the second regime when $\gamma \simeq \gamma_c$ and $\sigma\simeq\sigma_c$ (see Movie~1 in the Supplemental Material). We predict the viscoelastic stress response using a time-strain separable equation of the integral K-BKZ type \cite{Bird:1987,Larson:1988}. The stress is given by (see ref. \cite{Jaishankar:2014} for a more detailed derivation):\vspace{-2.0mm}
\begin{equation}
\sigma (t) = \int_{-\infty}^{t}G(t-t'){h}(\gamma)\dot \gamma (t'){\rm d}t'\label{eq0}
\end{equation} 
The first regime in Fig.~\ref{fig3} is fully accounted for by the linear viscoelastic response based on the power-law behavior of $G(t)$ determined in Fig.~\ref{fig1} for $\gamma_0\leq 1$\%. Since $h(\gamma)=1$ in this regime the stress can be found analytically \cite{Tanner:1988,Venkataraman1990}:\vspace{-2.0mm}
\begin{equation}
\sigma (t) = \int_{-\infty}^t G(t-t')\dot \gamma(t'){\rm d}t'=\frac{\mathds{V}\dot \gamma_0 t^{1-\alpha}}{(1-\alpha)\Gamma(1-\alpha)}\,. \label{eq1}
\end{equation}    
Equation~(\ref{eq1}) nicely describes the experimental data for $\gamma\lesssim 0.2$ without any additional fitting parameter [see dashed gray line in Fig.~\ref{fig3}(a) and (c)]. To predict the nonlinear behavior, we substitute the power-law form of the relaxation modulus $G(t)$ into Eq.~(\ref{eq0}) \cite{Rolon:2009,Mitsoulis:2013,Jaishankar:2014} and rearrange to give:\vspace{-2.0mm}
\begin{equation}
\sigma (t) = \frac{\mathds{V}\dot \gamma_0^\alpha}{\Gamma(1-\alpha)} \int_0^{\dot \gamma_0 t} {h}(\gamma)\gamma^{-\alpha}{\rm d}\gamma \, \label{eq2}
\end{equation}
where $\gamma=\dot{\gamma}_0t$ is the total accumulated strain at time $t$.  
To capture the strain-stiffening in the hydrogel we substitute the strain hardening form of the damping function for a protein gel, $h^*(\gamma)$, into Eq.~(\ref{eq2}) to obtain the prediction shown by the solid black line in Fig.~\ref{fig3}(a) and (c). This captures the nonlinear response of the gel at moderate strains, but leads to an ever-increasing rate of stress growth. The softening part of the damping function is crucial to account for the stress evolution observed experimentally during shear start-up.  
Substituting $\tilde{h}$ in Eq.~(\ref{eq2}) and integrating numerically we get the red line in Fig.~\ref{fig3}(a) and (c) which accurately predicts the mechanical response of the protein gel up to the stress maximum, without any adjustable parameter. The initial stiffening behavior is described by the numerator of $\tilde h$, while the denominator is responsible for the plateauing of the predicted stress response. The subsequent decrease of the stress observed experimentally in Fig.~\ref{fig3} must be associated with the growth of macroscopic fractures that cannot be accounted for by an integral formulation such as the K-BKZ equation, for which the value of the integral always increases monotonically in time.   
Repeating shear start-up experiments for various $\dot \gamma_0$ confirms that Eq.~(\ref{eq2}) quantitatively predicts the gel response over almost three decades of shear rate [Fig.~\ref{fig3}(b)]. The universal nature of the response is evident by the rescaling of the experimental data onto a single master curve [Fig.~\ref{fig3}(c)] and this rescaling also holds true for the 8\% wt gel [see Fig.~2(a) in the Supplemental Material].  

\begin{figure}[t!]
\centering
\includegraphics[trim={0 120 0 130},clip,width=1\columnwidth]{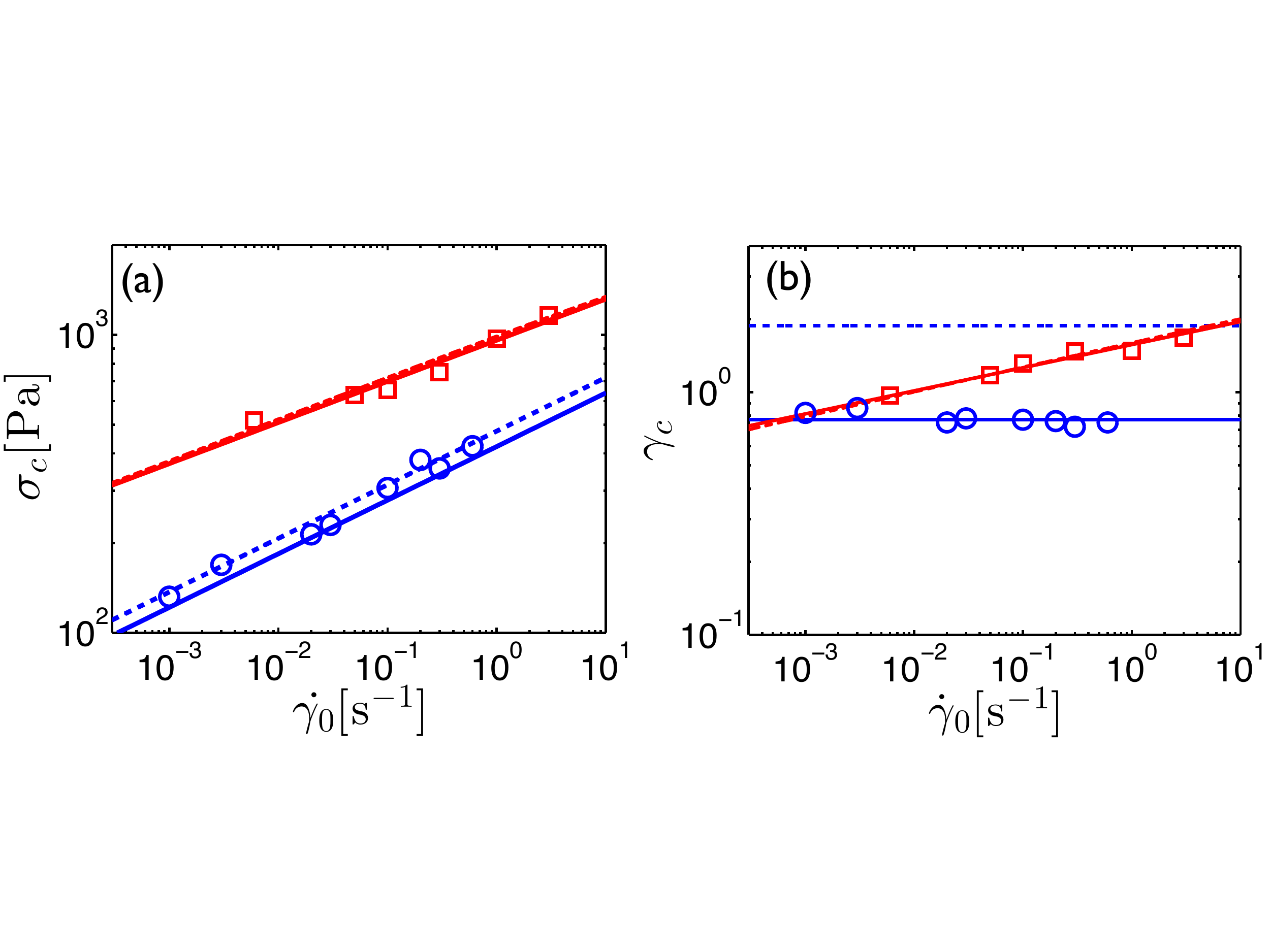}\vspace{-2mm}
\caption{(color online) Critical stress $\sigma_c$ (a) and critical strain $\gamma_c$ (b) vs applied shear rate $\dot \gamma_0$ for both a 4\% ($\circ$)  and a 8\% ($\square$) casein gel. In both graphs, the dashed and continuous lines stand for the prediction issued from the combination of the Bailey criterion and the stress response computed either from just the linear response (dashed lines) or the full K-BKZ equation built upon $h^*$ (solid lines). 
\label{fig4}\vspace{-3mm}}
\end{figure}

Moreover, the steady-state stress value predicted by the K-BKZ formulation coincides closely with the value of the stress maximum observed experimentally [Fig.~\ref{fig3}(b)], suggesting that there is a deeper connection between the failure point of the gel characterized by $\sigma_c$ and $\gamma_c$, and the nonlinear viscoelastic response of the material preceding macroscopic failure.
In Fig.~\ref{fig4} we show the locus of the stress maximum ($\sigma_c,\gamma_c$) for different imposed shear rates, for both the 4\% wt. and 8\% wt. casein gels. The critical stress $\sigma_c$ (above which macroscopic fractures appear) increases as a weak power law of $\dot \gamma_0$ with an exponent $\xi=0.18\pm 0.01$ for the 4\% wt. casein gel and $\xi=0.13\pm 0.01$ for the 8\% wt. respectively [Fig.~\ref{fig4}(a)]. Moreover, the critical strain $\gamma_c$ also displays a power-law increase with $\dot{\gamma_0}$ for the 8\% casein gel, whereas the 4\% casein gel shows a yield strain that is rate independent [Fig.~\ref{fig4}(b)]. The goal is now to predict such nontrivial power-law dependences.

\textit{Failure criteria and discussion.-} To account quantitatively for the different scalings of the crack appearance coordinates $\sigma_c$ and $\gamma_c$ with $\dot \gamma_0$, we apply the failure criterion introduced by J.~Bailey, already successful to describe the rupture of much stiffer samples such as glasses \cite{Bailey:1939} and elastomeric-like materials \cite{Malkin:1997}. This criterion may be applied under the assumption that the failure process is irreversible and results from independent damage events \cite{Freed:2002}. This appears to be the case for the brittle-like failure scenario of casein gels that are well modeled by Fiber Bundle Models that verify the former hypothesis \cite{Leocmach:2014,Kun:2008,Halaz:2012}. 
The Bailey criterion reads $\int_0^{\tau_f} {\rm d}t/F[\sigma(t)]=1$, where $\tau_f$ denotes the sample lifespan under an arbitrarily given active loading process $\sigma(t)$, and $F(\sigma)$ is the dependence of the time to rupture for creep experiments performed at a series of constant imposed stresses $\sigma$ \cite{Bartenev:1968}. 
Independent creep tests have been performed on casein gels \cite{Leocmach:2014} and indicate that $F(\sigma) = A \sigma^{-\beta}$ where $A$ is a scale parameter, with $A=(7.6\pm 0.1)\times10^{13}$~s.Pa$^{\beta}$ and $\beta =5.5\pm0.1$ for the 4\% wt. gel \cite{Leocmach:2014} and $A=(5.0\pm 0.1)\times10^{18}$~s.Pa$^{\beta}$ and $\beta =6.4\pm 0.1$ for the  8\%~wt.  gel [see Fig.~3 in the Supplemental Material]. We have also independently determined the rheological response to an arbitrary loading history [Eq.~(\ref{eq1}) and ~(\ref{eq2})]. When combined with Eq.~(\ref{eq1}), the Bailey criterion leads to the following analytic expressions for the critical strain and stress at failure under startup of steady shear: 
\begin{equation}
\gamma_c(\dot \gamma_0) = S_{\gamma} \dot \gamma_0^{(1- \alpha\beta)/[1+(1-\alpha)\beta]}\, \label{eq4}
\end{equation}
\vspace{-8mm}
\begin{equation}
\sigma_c(\dot \gamma_0) = S_{\sigma} \dot \gamma_0^{1/[1+(1-\alpha)\beta]}\, \label{eq3}
\end{equation}
where $S_{\sigma}$ and $S_{\gamma}$ are analytic functions of $\alpha$, $\mathds{V}$ and $\beta$ \cite{remark4}. Whether the critical stress and strain are constant or increase/decrease with $\dot{\gamma_0}$ thus depends on both the parameters in the linear viscoelastic kernel $G(t)$ and on the form of the failure law $F(\sigma)$. For example, in the 8\% wt. casein gel (for which $\alpha=0.04\pm0.01$ and $\beta=6.4\pm0.1$) we find that the critical strain increases with $\dot{\gamma_0}$ since the exponent in Eq.~(\ref{eq4}) is $(1-\alpha\beta)/\left[1+(1-\alpha)\beta\right]\simeq 0.10\pm0.01$. For both the 4\% wt. and 8\% wt. gels, the agreement between theory (dashed lines in Fig.~\ref{fig4}) and experiments is excellent for the two power-law exponents, again without any adjustable parameter. However, the prefactor $S_{\gamma}$ is clearly overestimated for the 4\% wt. casein gels (dashed line in Fig.~\ref{fig4}). Indeed, 4\% wt. casein gels display a pronounced stiffening responsible for the early rupture of the gel, which is not captured by the linear viscoelastic formulation Eq.~(\ref{eq1}). Instead, when combined with the Bailey criterion, direct numerical integration of Eq.~(\ref{eq2}) accounting for strain-stiffening leads to the correct value of the prefactor for both 4\% and 8\% casein gels (solid lines in Fig.~\ref{fig4}). Hence, one should use the complete damping function $\tilde{h}$ for predicting the failure point of strain-hardening materials. \\ 
\textit{Conclusion.-} We have shown using soft viscoelastic casein gels that the combination of the K-BKZ formalism, together with Bailey's failure criterion sets a self-consistent framework to capture the linear and non-linear response of the gel up to the materials' yield point. This approach also predicts the scaling of both the critical stress and strain at failure. The present results thus extend the validity of Bailey's criterion to squishy soft solids, paving the way for deeper analogies between soft and hard materials \cite{Schmoller:2013}. 

\begin{acknowledgments}
The authors thank M.~Bouzid and E.~Del Gado for enlightening discussions, and A. Parker for providing the casein and GDL. This work was funded by the MIT-France seed fund and by the CNRS PICS-USA scheme (\#36939). BK acknowledges financial support from Axalta Coating Systems.      
\end{acknowledgments}

\setcounter{figure}{0}
\clearpage
\onecolumngrid
\section*{\large{Nonlinear viscoelasticity and generalized failure criterion for polymer gels: supplemental material}}
\twocolumngrid

\begin{figure}[t!]
\centering
\includegraphics[width=0.8\columnwidth]{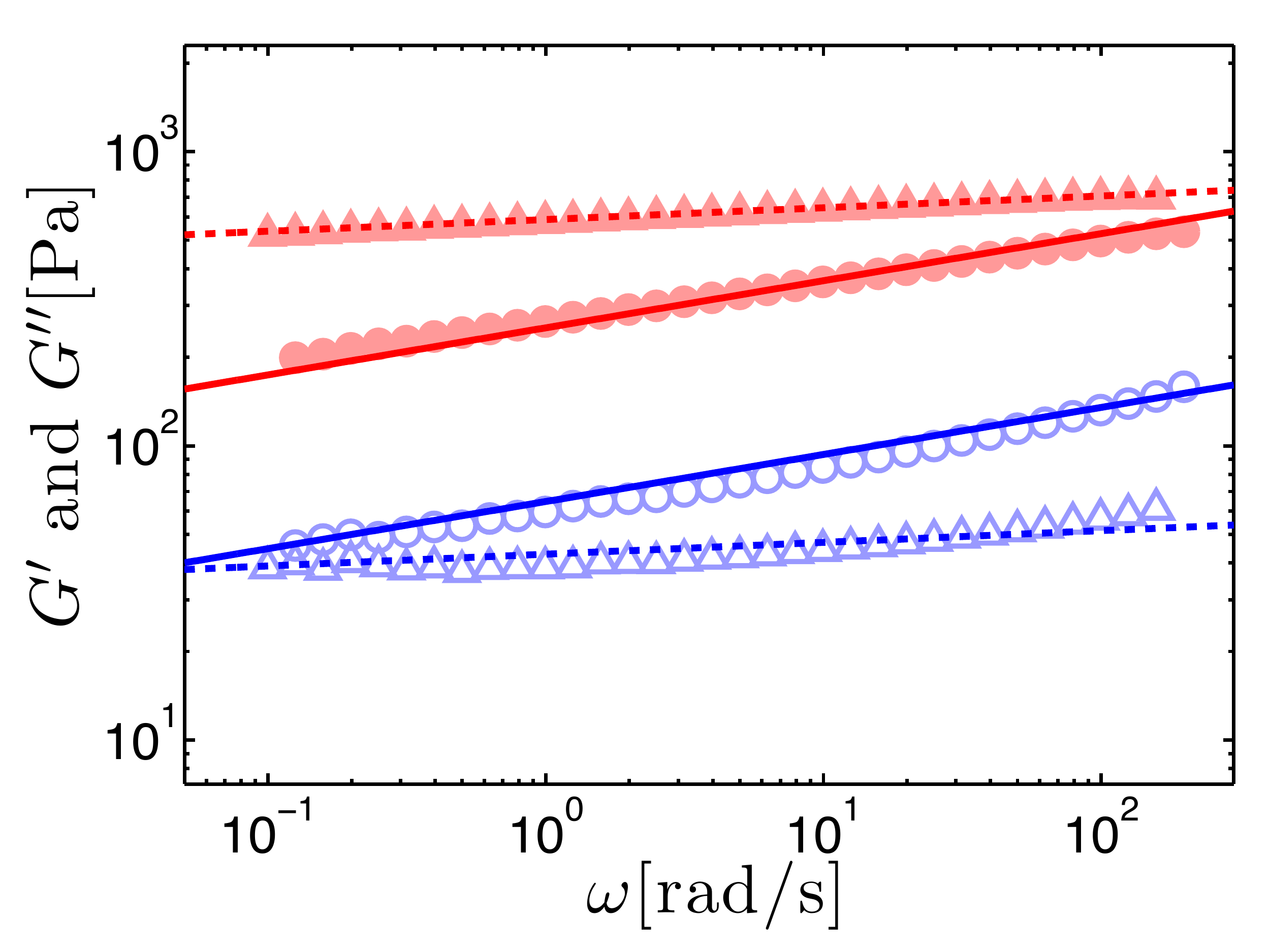}
\caption{(color online) Linear viscoelastic moduli $G'$ (upper, filled symbols) and $G''$ (lower, hollow symbols) as a function of pulsation $\omega$ for a strain amplitude of $\gamma=0.01$. The symbols $\bullet$ and $\blacktriangle$ respectively correspond to the 4\%~wt. and 8\%~wt. casein gels. Dashed and continuous lines correspond to the best fit of the data with the spring-pot model.
\label{suppfig1}}
\end{figure} 

\subsection*{Supplemental movie}

Supplemental Movie~1 shows the failure of a 4\%~wt. casein gel acidified with 1\%~wt. GDL in a Couette geometry for an imposed shear rate $\dot \gamma=$10$^{-3}$~s$^{-1}$. The rheological response recorded simultaneously corresponds to that shown in Fig.~3 in the main text. The first two cracks nucleate simultaneously at the top and bottom of the inner cylinder with slightly different angular positions and grow towards each other, stopping at the center of the cell. Meanwhile, a second pair of cracks nucleates next to the first one and grows in the same fashion before a third pair develops, and so on. Such a failure front propagates along the cell perimeter while the stress decreases in the third regime.

\begin{figure}[!t]
\centering
\includegraphics[width=0.9\columnwidth]{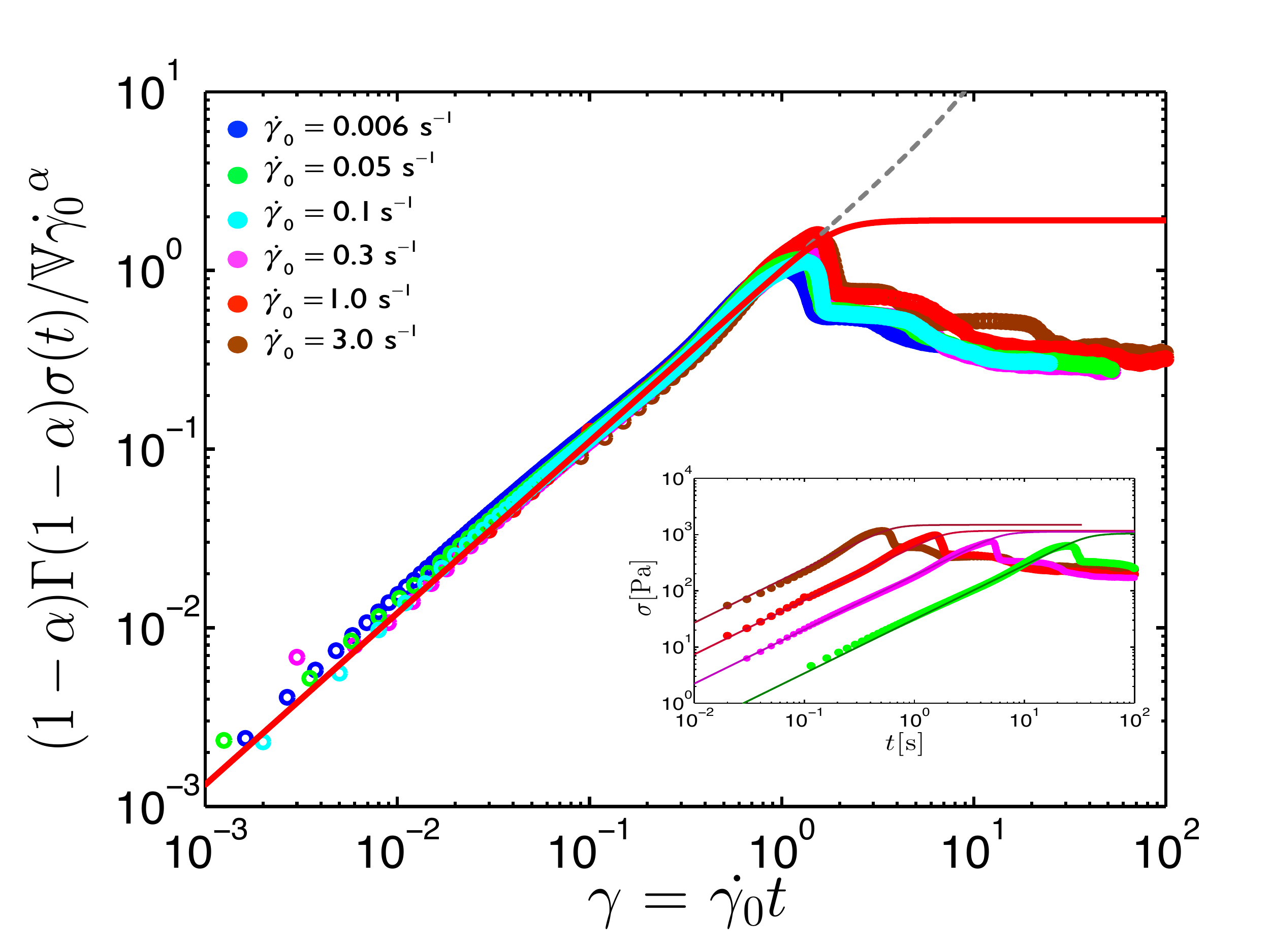}
\caption{(color online) Normalized stress responses $(1-\alpha)\Gamma(1-\alpha)\sigma (t)/\mathds{V}\dot \gamma_0^{\alpha}$ vs strain $\gamma=\dot \gamma_0 t$ in the start up of steady shear flow experiments for shear rates ranging from 0.006~s$^{-1}$ to 3~s$^{-1}$ for a 8\%~wt. casein gel. The dashed gray line corresponds to the linear response (Eqn. 2 in the main text) while the red solid line stands for the K-BKZ prediction constructed from the power-law linear viscoelastic response plus the nonlinear damping function $\tilde{h}$ determined independently from step strain experiment (see main text). Inset: non-normalized stress responses for a subset of shear start-up experiments shown in the main graph.
\label{suppfig2}}
\end{figure} 
\begin{figure}[!h]
\centering
\includegraphics[width=0.9\columnwidth]{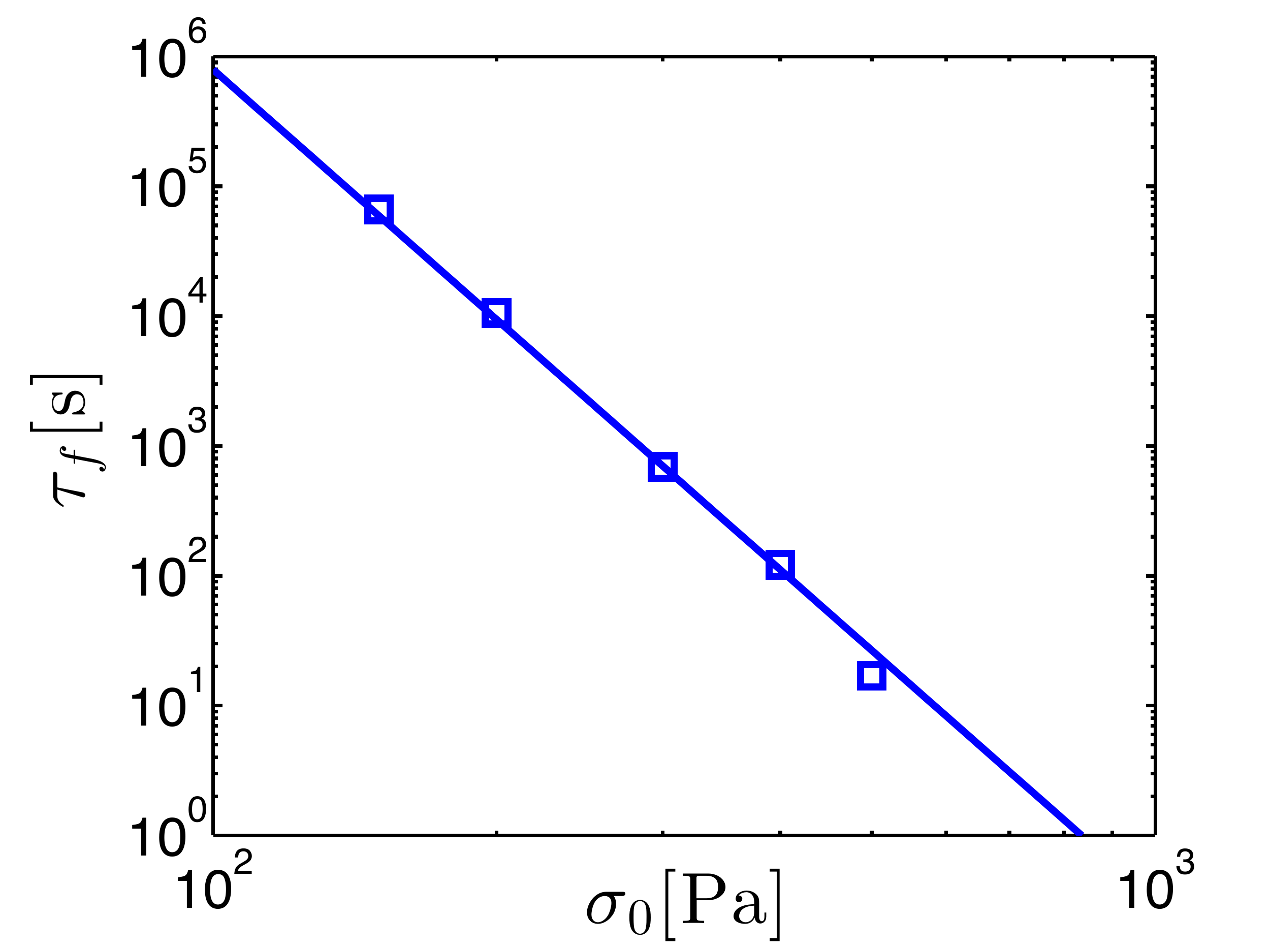}
\caption{(color online) Failure time $\tau_f$ as a function of the constant stress $\sigma_0$ applied during creep experiments. The solid line is the best power-law fit $\tau_f=A\sigma^{-\beta}$, with $A=(5.0\pm 0.1)\times10^{18}$~s.Pa$^{\beta}$ and $\beta =6.4\pm 0.1$ for the  8\% wt. gel. 
\label{suppfig3}}
\end{figure} 

\subsection*{Supplemental figures}

\textcolor{black}{As discussed in the main text we performed tests with two protein gels: $4\%$ casein-$1\%$ GDL and the $8\%$ casein-$8\%$ GDL gels which we henceforth refer to as  4\% and 8\% casein gels, respectively.}

Supplemental Figure~\ref{suppfig1} shows the frequency dependence of the elastic and viscous moduli of both the 4\%~wt. and 8\%~wt. gels. Both gels display a power-law linear rheology that can be modeled by a spring-pot (or fractional) element, as reported in the main text. The elastic and viscous modulus reads respectively $G'(\omega)=\mathds{V}\omega^\alpha \cos(\pi \alpha/2)$ and $G'(\omega)=\mathds{V}\omega^\alpha \sin(\pi \alpha/2)$, with $\alpha=0.18\pm 0.02$ and $\mathds{V}=261\pm 5$~Pa.s for the 4\% gel and $\alpha =0.04\pm 0.01$ and $\mathds{V}=620 \pm 5$ for the 8\% gel. 	 

Supplemental Figure~\ref{suppfig2} shows the stress responses to shear start-up experiments at different shear rates, which can be rescaled into a single master curve. The scaling of the failure time as a function of the applied stress, measured for creep experiments is plotted in Fig.~\ref{suppfig3}.


\begin{thebibliography}{}
\bibitem{remark0} Rheological data were obtained with an ARES-LS strain-controlled rheometer (TA Instrument) using a homemade polished Plexiglas rotating cup of height 20.32~mm and diameter 25~mm with an inner Delrin fixed bob of radius 22.65~mm, yielding a gap of 2.35~mm.
\bibitem{footnote} Therefore independent step strain experiments, instead of traditional strain sweeps which accumulate larger total deformations, allow us to prevent the growth of cracks and to accurately measure the initial stiffening and subsequent softening behavior of such polymer gels.
\bibitem{remark4} When $h(\lambda)=1$ the integrals may be evaluated analytically to give:\\$
S_{\sigma}=\left[\frac{A^{1-\alpha}\left[1+\beta\left(1-\alpha\right)\right]^{1-\alpha}\mathbb{V}}{(1-\alpha)\Gamma(1-\alpha)}\right]^{1/\left[1+\beta\left(1-\alpha\right)\right]}$ and\\ $S_{\gamma}=\left[A\left[1+\beta\left(1-\alpha\right)\right]\left(\frac{(1-\alpha)\Gamma(1-\alpha)}{\mathbb{V}}\right)^\beta\right]^{1/\left[1+\beta\left(1-\alpha\right)\right]}$
\end{thebibliography}
\end{document}